\documentclass[letterpaper,conference]{IEEEtran}

\usepackage[english]{babel}
\usepackage{tabularx}
\usepackage{xspace} % for fixing the space after newcommand
\usepackage{booktabs}
\usepackage{pgf}
\usepackage{subcaption}
\usepackage{graphicx}
\usepackage{amsmath}
\usepackage{amssymb}
\usepackage{fnpct} %footnotemark after "." 

\newtoggle{comments}
\toggletrue{comments}
\iftoggle{comments}{\newcommand{\TODO}[1]{{\color{red}{TODO: #1}}}}{\newcommand{\TODO}[1]{}}

\begin{document}

\title{Learning Wi-Fi Connection Loss Predictions for Seamless Vertical Handovers Using Multipath TCP}

\author{
	\IEEEauthorblockN{
		Jonas H\"{o}chst\IEEEauthorrefmark{1}\IEEEauthorrefmark{2}, 
		Artur Sterz\IEEEauthorrefmark{1}\IEEEauthorrefmark{2}, 
		Alexander Fr\"{o}mmgen\IEEEauthorrefmark{1}, 
		Denny Stohr\IEEEauthorrefmark{1}, 
		Ralf Steinmetz\IEEEauthorrefmark{1}, 
		Bernd Freisleben\IEEEauthorrefmark{1}\IEEEauthorrefmark{2}
	} \\
	\IEEEauthorblockA{
		\IEEEauthorrefmark{1}\textit{Dept. of Electrical Engineering \& Information Technology}, TU Darmstadt, Germany\\
		E-mail: \{jonas.hoechst, artur.sterz, alexander.froemmgen, denny.stohr, ralf.steinmetz, bernd.freisleben\}@maki.tu-darmstadt.de
	}
	\IEEEauthorblockA{
		\IEEEauthorrefmark{2}\textit{Dept. of Mathematics \& Computer Science}, Philipps-Universit\"{a}t Marburg, Germany \\
		E-mail: \{hoechst, sterz, freisleb\}@informatik.uni-marburg.de
	}
}

\maketitle

% add page numbers
% \thispagestyle{plain}
% \pagestyle{plain}

\begin{abstract}

We present a novel data-driven approach to perform smooth Wi-Fi/cellular handovers on smartphones.
Our approach relies on data provided by multiple smartphone sensors (e.g., Wi-Fi RSSI, acceleration, compass, step counter, air pressure) to predict Wi-Fi connection loss and uses Multipath TCP to dynamically switch between different connectivity modes. 
We train a random forest classifier and an artificial neural network on real-world sensor data collected by five smartphone users over a period of three months. 
The trained models are executed on smartphones to reliably predict Wi-Fi connection loss 15 seconds ahead of time, with a precision of up to 0.97 and a recall of up to 0.98.
Furthermore, we present results for four DASH video streaming experiments that run on a Nexus 5 smartphone using available Wi-Fi/cellular networks.
The neural network predictions for Wi-Fi connection loss are used to establish  MPTCP subflows on the cellular link.  
The experiments show that our approach provides seamless wireless connectivity, improves quality of experience of DASH video streaming, and requires less cellular data compared to handover approaches without Wi-Fi connection loss predictions.

\end{abstract}

\section{Introduction}
\label{sec:introduction}
Smartphones have become our daily mobile companions to provide wireless access to communication, information, and entertainment services. Since a large amount of data is not downloaded in advance but strea\-med on demand via the Internet, seamless connectivity using both Wi-Fi and cellular interfaces is desirable.
The mobility of smartphone users leads to the problem of deciding when to use which wireless connection. Most smartphones use Wi-Fi as their default interface, since many cellular data plans will be throttled after exceeding a certain limit. The decision when to perform vertical handovers is often based on the Wi-Fi received signal strength indicator (RSSI) and timeouts for the transmission of packets.
%~\cite{prevFoot}. 
However, if a user is leaving a location while listening to a music stream or watching a streamed video, the established Wi-Fi connection eventually becomes unavailable and the streaming process stops. The mobile operating system detects the connection loss some time after the connection is lost. Finally, the application needs to reestablish the connection. 

In this paper,  we present a novel approach to predict Wi-Fi connection loss before the connection breaks to perform seamless vertical Wi-Fi/cellular handovers. Our approach relies on data collected by multiple smartphone sensors (e.g., Wi-Fi RSSI, acceleration, compass, step counter, air pressure) to predict Wi-Fi connection loss and uses Multipath TCP (MPTCP) to dynamically switch between different wireless connectivity modes. 
We train a random forest classifier and an artificial neural network on roughly 20 GB of sensor data collected by five smartphone users over a period of three months. 
The trained models are efficiently executed on smartphones and reliably predict Wi-Fi connection loss 15 seconds ahead of time, with a precision of up to 0.97 and a recall of up to 0.98.
Furthermore, we present results of four DASH video streaming experiments that run on an Android smartphone and make use of available Wi-Fi/cellular networks.
The neural network predictions for Wi-Fi connection loss are used to establish  MPTCP subflows on the cellular link.  
Our experiments show that the proposed approach provides seamless wireless connectivity, improves quality of experience by increasing mean opinion scores (MOS) from 2.7 to up to 3.8 for certain scenarios, and requires up to 50\% less cellular data compared to handover approaches without Wi-Fi connection loss predictions.
%Furthermore, the power overhead is negligible with only 2.29\% additional power usage, which reduces the continuously watchable time on the smartphone by only six minutes.
The data set, analysis scripts, experimental logs, and the mobile app developed in this paper are publicly available\footnote{https://umr-ds.github.io/seamcon/}.
%are available at \url{https://seamcon.net}. 
%\TODO{Find a nicer available url?}
%\TODO{Make a decision which data to be released.}
%
To summarize, we present:

\begin{itemize}
\item a novel approach to predict Wi-Fi connection loss for performing seamless vertical handovers,
\item a neural network to learn and predict Wi-Fi connection loss based on a novel combination of smartphone sensors,
\item a vertical handover method that uses MPTCP for switching between wireless connection modes at runtime,
\item an implementation on off-the-shelf smartphones to demonstrate its performance in real-world scenarios; our results show significant improvements in terms of Quality of Experience and the amount of cellular data consumed. 
\end{itemize}

The  paper is organized as follows.
Section~\ref{sec:rel_work} discusses related work. Section \ref{sec:design} presents an overview of our approach. Section \ref{sec:eval} analyzes the methods for performing predictions, and Section \ref{sec:eval2} evaluates the performance using vertical Wi-Fi cellular handovers on smartphones.
%connection loss predictions on smartphones for vertical handovers based on MPTCP.
Section~\ref{sec:conclusion} concludes the paper and outlines areas for future work.

\section{Related Work}
\label{sec:rel_work}

\subsection{Predicting Wi-Fi Connection Loss}
Several approaches to predict Wi-Fi connection loss for performing handovers have been proposed in the literature ~\cite{6587998}.
Nasser et al. \cite{4289611} 
% and Bhattacharya~et~al.~\cite{4426586} 
use neural networks to predict Wi-Fi connection loss events based on RSSI.
Horich et al.~\cite{4138149} use a fuzzy logic controller (FLC) for making decisions about performing handovers, where the parameters for the FLC are learned using a neural network.

Lin et al.~\cite{Lin:2008:NCH:1386109.1386110} propose to use standard Wi-Fi connection properties and a neural network to predict Wi-Fi connection loss.
% packet length, RSSI, signal-to-noise ratio (SNR), the number of connected users, and a self-defined packet success rate indicator for training a neural network for Wi-Fi connection loss predictions. 
% Wang et al.~\cite{WANG20101122} predict Wi-Fi connection loss based on link quality, taking into account device configurations and network conditions.
Monsour et al.~\cite{Mansour:2017:SHB:3090354.3090423} use a combination of user velocity and the Allan variance of the RSSI to predict Wi-Fi connection loss, and use PMIPV6 to manage the predicted Wi-Fi connection loss.
Khan et al.~\cite{Khan2017} propose a fuzzy logic system to predict Wi-Fi loss events based on various parameters, such as delay, jitter, bit error rate, packet loss, communication cost, response time, and network load.

These approaches are limited to information of wireless connections, which may be helpful to create metrics for Wi-Fi quality, but is not always the best information for predicting Wi-Fi connection loss. In contrast, our approach considers information from a wide range of smartphone sensors that indicate the usage context, leading to high-quality predictions.

% würde hier nochmal einen schlusssatz reinbringen, der die abgrenzung bringt.
% erst offensichtlichen punkt, dann noch den punkt, dass es damit eingeschränkt ist auf fälle, bei denen die Verbindung auch tatsächlich genau in dem zeitraum genutzt wird. 

% Hayes et al.~\cite{7158091} consider historical network performance, current network characteristics, and the current weather situation to predict Wi-Fi connection loss.

Other approaches incorporate
the mobility of the users~\cite{Nicholson:2008:BFM:1409944.1409952}
% the GSM location area, surrounding Bluetooth devices and Wi-Fi access points, charging state and wired network connectivity~\cite{Rathnayake:2009:DAN:1641804.1641836},
or higher level features like social group affiliation, time-of-day, and average duration a user spends in a particular network~\cite{Wanalertlak:2011:BMP:1969330.1969339}.
% Wanalertlak et al.~\cite{Wanalertlak:2011:BMP:1969330.1969339} incorporate location and mobility traces of users, social group affiliation, time-of-day, and average duration a user spends in a particular network.
% Rathnayake et al.~\cite{Rathnayake:2009:DAN:1641804.1641836} use a dynamic Bayesian network model based on GSM location area, number of Bluetooth devices around, charging state of the device, available Wi-Fi access points (APs), time-of-day, and LAN cable state.
% Nicholson and Noble \cite{Nicholson:2008:BFM:1409944.1409952} exploit the mobility of a user to forecast Wi-Fi conditions. The authors combine the recorded user traces with TCP speed test data and a Wi-Fi database for their predictions of Wi-Fi connection loss.
%
% Geurts et al.~\cite{10.1007/978-3-319-16486-1_93} use a crowdsourced connectivity map to forecast the available bandwidth in the near future to prefetch data. 
% Depending on the information gathered from this map, their system prefetches data in good conditions to have a sufficient buffer during bad connections or handover situations.

The predictions in all of these approaches depend on external factors and indicators. In contrast, our approach only requires information that every current mobile device provides and thus can be used in a straightforward, economically attractive manner. To best of our knowledge, there is no work that uses smartphone sensor data to predict  Wi-Fi connection loss. 

\subsection{Performing Vertical Handovers}

There are extensions to the traditional Internet Protocol that allow users to keep a session alive when (vertical) handovers are performed \cite{rfc5944}. These approaches are based on home and foreign agents that forward traffic for the mobile host. Although they are around for a long time, mobile IP is not supported widely.
Ma et al.~\cite{ma2004new} propose a vertical handover method based on the Stream Control Transmission Protocol. While the proposed method is network-independent and thus does not require home and foreign agents, traditional TCP-based applications cannot benefit from the advancements. 
MPTCP is a TCP extension supporting multiple subflows for a single TCP connection~\cite{rfc6824}.
MPTCP improves throughput and reliability in data center and mobile environments~\cite{raiciu2012hard, chen2013measurement}.
% Han et al.~\cite{Han:2016:MAV:2999572.2999606} propose preference-aware video streaming over MPTCP by using cellular connections if the Wi-Fi throughput is not sufficient.
%
Paasch et al.~\cite{PDDRB12} evaluate MPTCP as a vertical handover mechanism. The authors propose three MPTCP modes for handover scenarios, namely \textit{Full}, \textit{Backup}, and \textit{Single-Path Mode}. The first two modes maintain subflows on all interfaces, while the \textit{Single-Path Mode} exploits the break-before-make design of MPTCP.
Pluntke et al.~\cite{pluntke2011saving} use MPTCP as a vertical handover mechanism to shift connections between cellular and WiFi connectivity and finally to save energy.
De Coninck and Bonaventure~\cite{de2017every} futher improve the handover by speeding up packet retransmissions after the cellular subflow is established.

The handover mechanisms in these approaches are either reactive, resulting in temporary connection losses, or use redundancy, leading to high bandwidth consumption, which is often contrary to the users' preferences.

% \subsection{Predicting Disk and DRAM Replacement}
% Giurgiu et~al.~\cite{giurgiu2017predicting} use random forests to predict DRAM reliability.
% They model their problem as shown in \ref{fig:dram_paper}.

% Botezatu et~al.~\cite{Botezatu:2016:PDR:2939672.2939699} use SMART data from hard disks to predict disk reliability with a \emph{greedy forest}.

% \textbf{Discussion}
% In contrast to their work, our timing is a bit more important.
% It is sufficient to forecast that the module will fail in the next two weeks.

% \begin{figure}[t]
%     \centering
%     \includegraphics[width=\linewidth]{figures/model_from_dram_paper.pdf}
%     \caption{Model from DRAM paper.}
%     \label{fig:dram_paper}
% \end{figure}
\section{Conceptual Overview}
\label{sec:design}
\begin{figure}[t]
    \centering
    \includegraphics[width=\linewidth]{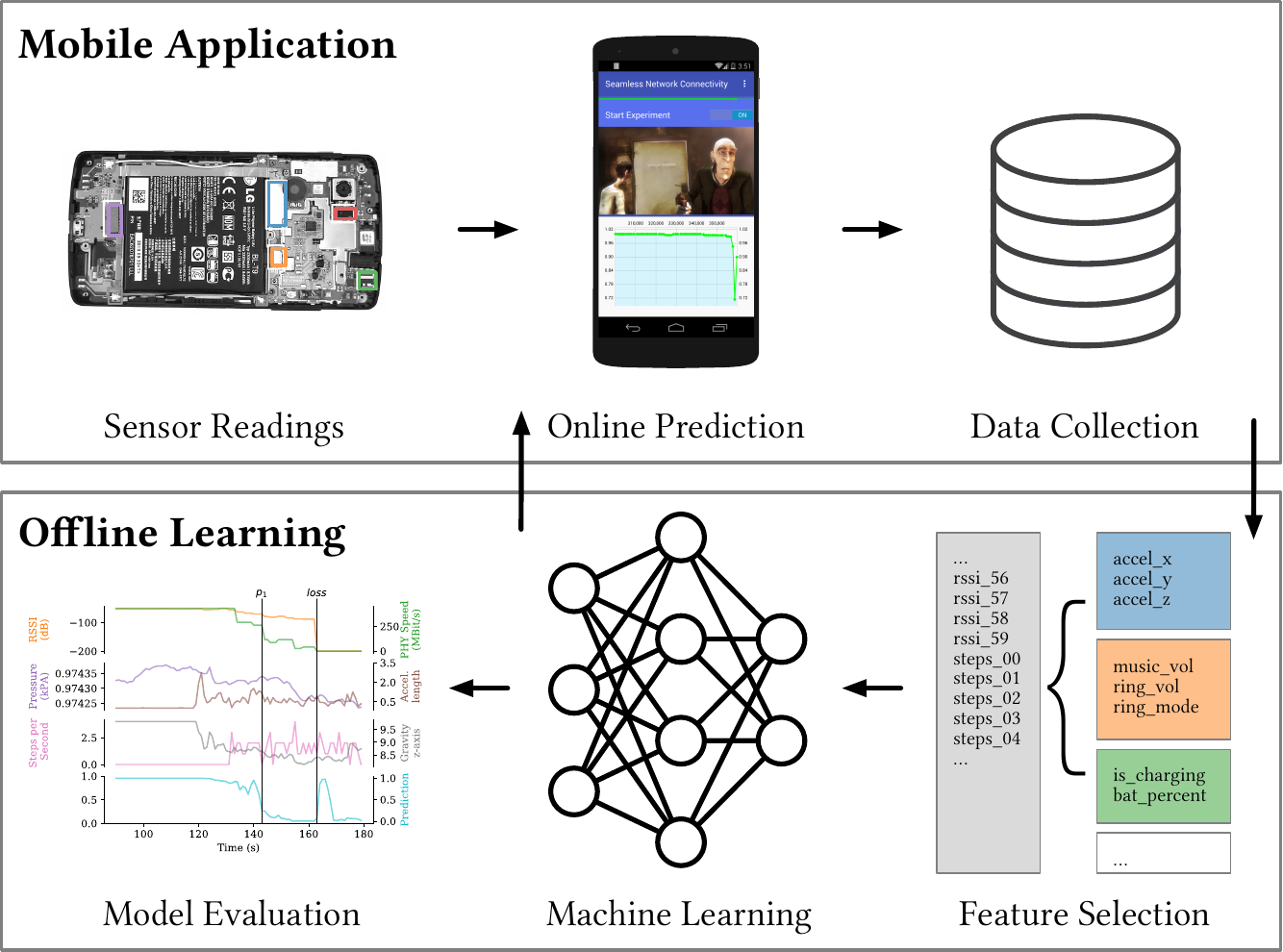}
    \caption{Mobile application and offline learning.}
    \label{fig:learning_workflow}
\end{figure}

Figure~\ref{fig:learning_workflow} shows the components of our approach and the workflow. First, raw sensor data is collected by a mobile app developed for our work and uploaded to a server for further processing.  The raw data is appropriately preprocessed and enriched with additional higher level features. The resulting data is then used to train and evaluate a random forest classifier and different neural network architectures. The data preprocessing operations as well as the trained models are transpiled to Java code and integrated into the mobile app on the smartphone, which in turn makes online predictions for Wi-Fi connection loss 15 seconds ahead of time.
% while the user moves around.
Based on these predictions, vertical Wi-Fi/cellular handover is performed using MPTCP. We explain the main steps of our approach in more detail below. Neural network model building and Wi-Fi connection loss predictions for performing MPTCP handovers on a smartphone are discussed in Sections \ref{sec:eval} and \ref{sec:eval2}, respectively.

\subsection{Smartphone Sensors}
\label{design:sensors}
Modern smartphones offer a variety of sensors that directly or indirectly measure different properties, as explained below. Even though every individual feature might not be a good {Wi-Fi} connection loss indicator, combinations of seemingly irrelevant features can improve the prediction accuracy. 

\paragraph*{Motion}
Depending on the abstraction level, direct motion sensor readings (e.g., accelerometer, gyroscope, magnetometer), sensor readings cleaned from unwanted influences  (e.g., gravity, linear acceleration, rotation vector), or higher level sensor readings as hardware processed triggers (e.g., significant motion, step counter, step detection), are good predictors for user movement.
% Motion sensors can be distinguished by their abstraction level:
% \begin{enumerate}
%     \item physical sensors offering direct sensor readings, such as accelerometer, gyroscope, or magnetometer;
%     \item sensor values preprocessed on a device that are cleaned from unwanted influences, such as  gravity, linear acceleration, or rotation vector;
%     \item higher level sensor readings that offer hardware processed triggers, such as significant motion, step counter, and step detection.
% \end{enumerate}

\paragraph*{Orientation}
Orientation sensors can reveal more specific situations, where a phone is in the pocket or laying on a table.
The proximity sensor is typically used to detect whether the smartphone is held to the ear, but can also be a good hint for other situations, e.g., to detect whether the smartphone is face down on the table.

% Orientation sensors are typically used for navigation. Orientation and current rotation are measured by either the geomagnetic field or a gyroscope sensor. 
% In addition, most smartphones make use of a proximity sensor to detect whether the smartphone is held to the ear during phone calls. While only available in one axis, this sensor can be a good hint for other situations, e.g., to detect whether the smartphone is in the pocket.

\paragraph*{Environment}
Environmental sensors include sensors for measuring ambient light to control screen brightness, humidity, air pressure, and ambient temperature. 
% Although these sensors might not be available on legacy smartphones, most modern devices contain them. 
Rapid changes in these sensor readings can reveal a sudden change of the smartphone situation, e.g., going outdoors.

\paragraph*{Global position}
GPS can be useful in combination with a world map of Wi-Fi availability. 
%Also when training for one specific user, a local map can be created. 
% The global positioning system (GPS) provides information about the smartphone's position and height. To reduce the time to obtain position information, it is often used in combination with a heatmap of Wi-Fi availability to determine the smartphone's position.
Due to quality concerns with indoor GPS traces and high energy consumption, we discarded GPS in our work.

\paragraph*{User interactivity}
The user's current context can be derived from various indicators, like device state (interactive, idle, power save), current charging state, audio state (speaker, headphones, and their volumes), and ringer mode.

\paragraph*{Wi-Fi properties}
Wi-Fi properties, obtained by the radio interface, provide insights into the current connection quality along with reachable networks. Relevant indicators include RSSI, data link layer speed, and used frequency bands. 
%Some sensors like SNR or  PHY mode were not considered because of their limited availability in older Android versions.

\subsection{Sensor Data Preprocessing}
\label{ssec:sensor_preprocessing}
To learn Wi-Fi connection loss predictions, the sensor data needs to be preprocessed. The time component of the sensor readings needs to be incorporated in the feature vector. 
%Similar problems have been solved using traditional machine learning approaches whilst encoding the time component through sliding windows for data input and classification. 
%Giurgiu et~al.~\cite{giurgiu2017predicting} use random forests to predict DRAM reliability using this pattern. Botezatu et~al.~\cite{Botezatu:2016:PDR:2939672.2939699} use the observation and prediction window approach for hard disk reliability in data centers in combination with greedy forests.
% We define the observation window $OW$ as the time frame in which the sensor readings are used as the inputs to the trained predictor, and the prediction window $PW$ as the time frame the classifier is trained on to predict ahead of time.
% Figure~\ref{fig:data_flow} provides an overview of our data preprocessing steps. The raw sensor readings are resampled to common timestamps, the sensor readings are enriched with higher level feature computations, and the observation window is created. The resulting feature vector is then used to train the neural network. These steps are explained in the following.

% \begin{figure}[t!]
%     \centering
%     \includegraphics[width=.96\linewidth]{figures/data_flow}
%     \vspace{-2mm}
%     \caption{From sensor readings to predictions.}
%     \label{fig:data_flow}
% \end{figure}

\paragraph*{Sensor sampling}
\label{design:sampling}
The used heterogeneous sensors have different reading frequencies. Motion and orientation sensors can be read with a rather high sampling rate $R$ of 50 Hz, while other sensors are available and useful just under 1 Hz. As a trade-off between energy consumption and sensor data quality, we chose a sensor data sampling rate of $R=1$ sample per second. Sensors with lower sampling rates are filled until a new value becomes available. 
% Implementation changed after CoNEXT review
% For sensors offering higher sampling rates, the average value within one second is used. 

% \begin{figure}[t!]
%     \centering
%     \includegraphics[width=.90\columnwidth, trim={0.25cm 0.1cm 0.25cm 0.1cm}, clip]{figures/observation-prediction.pdf}
%     \caption{Observation and prediction window.}
%     \label{fig:observation-prediction}
% \end{figure}

\paragraph*{Observation and prediction window}
\label{design:observation}
To enrich the discrete sensor readings and to consider the temporal component, the sensor readings are processed in an observation window $OW$. 
% Figure \ref{fig:observation-prediction} provides an illustration of $OW$ and $PW$ and synthetic sensor readings for the example of \textit{linear acceleration}, \textit{step count}, and \textit{Wi-Fi connectivity}. 
We use an observation window of 60 seconds, which is derived from common walking speeds and Wi-Fi access point ranges.
% In Figure \ref{fig:observation-prediction}, the observation window goes from t=-59 to t=0, thus in this case the last 60 readings of each sensor would be available to the prediction algorithm. 
% \paragraph*{Prediction window}
The earlier a Wi-Fi connection loss is predicted, the more effective the transition between Wi-Fi/cellular is. To define an upper bound on the prediction window, the quality characteristics of the used network protocols are important. Transport protocols, such as TCP, use slow-start to avoid congestion. To compensate for this low-bandwidth start, an early prediction is useful. 
% Most transport protocols, such as TCP, use slow-start in the first phase of a connection to avoid congestion \cite{allman2009tcp}.  To compensate for this low-bandwidth phase, an early prediction is useful. In general, the quality of predictions is lower when longer prediction windows are used.
As a trade-off between performance and farsightedness, and to avoid long-running redundant MPTCP connections, which are energy and data plan consuming, we use a prediction window of up to 15 seconds.
% The prediction indicates whether Wi-Fi will be available for the entire next prediction window. In Figure \ref{fig:observation-prediction}, the prediction at $t=0$ covers the entire $PW$ of 15 seconds and indicates that Wi-Fi connectivity will be lost (in this example it is lost at $t=11$).

%\paragraph{Prediction Target}
%Depending on the handover mechanism, different time and prediction probability thresholds can be applied to fine-tune the quality/cost ratio. For example, a handover to cellular as the default link will result in higher energy consumption and therefore should happen as late as possible. When using Wi-Fi with cellular as a backup connection in a MPTCP setup, the established cellular subflow would be less energy consuming and could happen more early. The large prediction window of 15 seconds allows a general approach for the learning phase, as well fine-tuning when using different transition strategies.

\paragraph*{Feature vector}
The feature vector presented to the learning algorithm consists of the sensor readings in the observation window. Each individual sensor contributes \mbox{$OW \times SR$} values to the feature vector. In the selected configuration, this results in 60 values per sensor.
Furthermore, all features are normalized by removing the mean and scaling to unit variance, as required for the machine learning algorithms used in our approach.

\subsection{Precision and Recall}
When the Wi-Fi connection loss is predicted too early or too often, this can result in higher consumption of the commonly restricted data plans of the users. Predicting it too late, on the other hand, can result in a dissatisfactory QoE. 
The primary goal is to reach a high recall in predicting Wi-Fi connection loss. In terms of energy efficiency, the secondary goal is to reach a high precision predicting Wi-Fi connection loss, thus not performing unnecessary handovers. 

% \newcommand{\userSven}{User 1\xspace}
% \newcommand{\userMsommer}{User 2\xspace}
% \newcommand{\userTimo}{User 3\xspace}
% \newcommand{\userLab}{User 4\xspace}
% \newcommand{\userAlex}{User 5\xspace}

% \begin{table}[t]
%     \centering
%     \caption{Data set collected for our work}
%     \label{dataset}
%     \begin{tabular}{lrrr}
%     \toprule
%     \textbf{User}  & \textbf{Size}    & \textbf{Proc. Elem.} & \textbf{Timespan} \\ \midrule
%     \userSven            & 9.7 GB           & 357 k                & 16 days              \\ 
%     \userMsommer         & 4.3 GB           & 328 k                & 28 days              \\ 
%     \userTimo            & 3.4 GB           & 140 k                & 75 days              \\ 
%     \userLab             & 0.5 GB           & 3 k                  & 2 days               \\ 
%     \userAlex            & 0.7 GB           & 81 k                 & 5 days               \\ \midrule
%     \textbf{$\sum$}      & \textbf{19.6 GB} & \textbf{909 k}       & \textbf{-}        \\ \bottomrule
%     \end{tabular}
% \end{table}

\section{Learning Wi-Fi Loss Predictions}
\label{sec:eval}
In this section, our novel data-driven approach to predict Wi-Fi connection loss is presented. %First, the data collected for this paper will be presented. The feature vector preparation and two sets of features considered will be presented. Finally a data example and the machine learning results will be discussed.

\subsection{Data Set}
We collected about 20 GB of smartphone sensor data from 5 users, with more than 900,000 unique samples, over a period of three months. The users were advised to let the mobile application run throughout the day, thus the traces contain data from the users' daily lives.
% Table \ref{dataset} provides an overview of the collected data.
% The term processed elements in the third column refers to the preprocessing step described in Section \ref{ssec:sensor_preprocessing}, and therefore only samples are used where Wi-Fi was available. Furthermore, all samples contain values for all features. If features were missing, the regarding samples were discarded.

\paragraph*{Training and test set}
Machine learning methods require separate data sets for training and testing to verify the generalization abilities of the trained models. We investigated different ways of building training and test sets: (a) we randomly split the available samples into, e.g., 70\% training and 30\% test data, and (b) we split by users, to learn and test with different users.
% e.g.,  \userSven, \userMsommer or \userTimo as the training set and use \userLab and \userAlex as the test set. Option (b) can later be improved through online learning using the data collected by the device's owner.

\subsection{Feature Vectors}
All features collected on the smartphones can be used as predictors for Wi-Fi connection loss. We used two feature vector sets, namely the \textit{Full} and the \textit{Reduced Feature Vector}.

\paragraph*{Full Feature Vector}
The data collected by the different users shows that some features are not available on all devices. The 25 features selected for the full feature vector consist of values of all available sensors:
% \begin{itemize}
%     \item Atmospheric pressure: $x$, $delta$,
%     \item Linear acceleration: $x$, $y$, $z$, $length$,
%     \item Step counter: $delta$,
%     \item Power: $is~charging$, $battery~percentage$,
%     \item Gravity: $x$, $y$, $z$,
%     \item Gyroscope: $length$,
%     \item Magnetic field: $x$, $y$, $z$,
%     \item Orientation: $x$, $y$, $z$,
%     \item Rotation: $x$, $y$, $z$,
%     \item Wi-Fi: $frequency$, $speed$, $RSSI$.
% \end{itemize}
Atmospheric pressure: $x$, $delta$; Linear acceleration: $x$, $y$, $z$, $length$; Step counter: $delta$; Power: $is~charging$, $battery~percentage$; Gravity: $x$, $y$, $z$; Gyroscope: $length$; Magnetic field: $x$, $y$, $z$; Orientation: $x$, $y$, $z$; Rotation: $x$, $y$, $z$; Wi-Fi: $frequency$, $speed$, $RSSI$.
Thus, the feature vector consists of $25 \times 60 = 1500$ features (i.e., with a 60 seconds observation window).

\paragraph*{Reduced Feature Vector}
Many of the sensors, like linear acceleration and gyroscope, described in Section~\ref{design:sensors} share underlying features due to their physical properties. The number of sensors can be reduced by leaving aside these sensors. For the \textit{Reduced Feature Vector}, we used the following sensors:
% \begin{itemize}
%     \item Atmospheric pressure: $delta$,
%     \item Linear acceleration: $length$,
%     \item Step counter: $delta$,
%     \item Power: $is~charging$,
%     \item Gravity: $z$,
%     \item Wi-Fi: $frequency$, $speed$, $RSSI$.
% \end{itemize}
Atmospheric pressure: $delta$; Linear acceleration: $length$; Step counter: $delta$; Power: $is~charging$; Gravity: $z$; Wi-Fi: $frequency$, $speed$, $RSSI$.

\subsection{Sensor Data Example}

% Alex: Ìrgendwo in der intro als intuitives beispiel bringen, dass man sich meistens bewegt, wenn man wifi verliert, und gerade die schrittsensoren so billig bzgl. strom sind, dass sie ohnehin bei den meisten geräten im Hintergrund ausgeführt werden... (idee dabei sit, dass wir wohl keine saubere quelle für den strom aspekt haben, also nehmen wir den sensor, von dem wir es am einfachsten behaupten können, ggf. sogar ne fußnote dazu, dass das meistverkaufte android handy so ein ding hat)

% ### Beginn Beispiel (verschoben aus Section 4.3)

\begin{figure}[t]
    % mlp_random_rssi_learner, databases["/shared/seamless-upload/upload/timo/1518785457.db"]
    \centering
    \includegraphics[width=\linewidth]{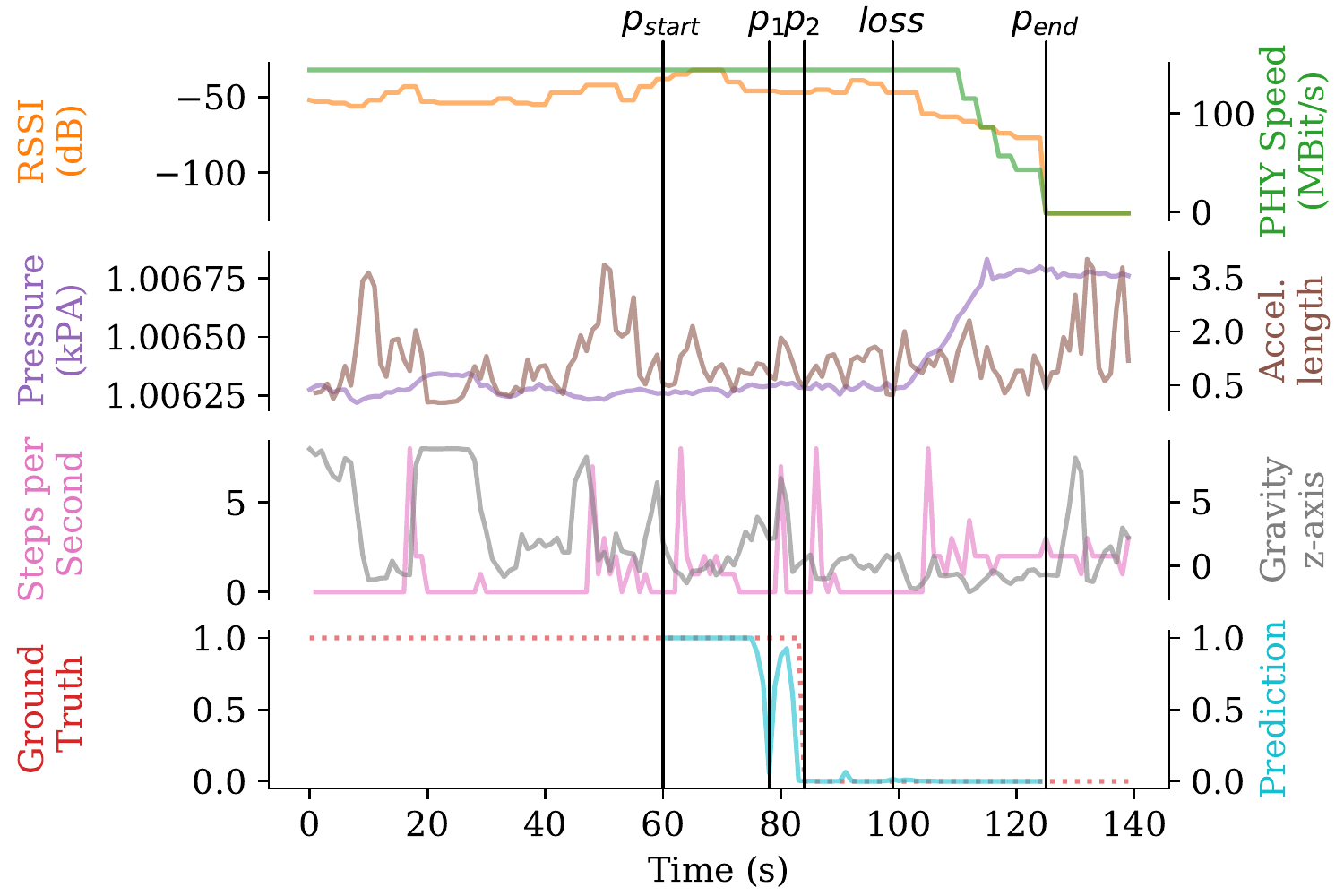}
    \caption{Different sensors leading to an early ($p_1$) and an ideal ($p_2$) prediction of Wi-Fi connection loss, based on a trained model with randomly split data.}
    % \TODO{Alex:wuerde hier nen halbsatz dazu sagen, was man erkennen soll... der leser ist faul, was sind die p? platz hier ist glaub ich gut angelegt.}}
    \label{fig:timo_1518785457}
\end{figure}

Figure~\ref{fig:timo_1518785457} shows an example of several sensor data values collected by a smartphone. 
% From top to bottom, the figure shows the Wi-Fi sensors RSSI and the current PHY speed,   atmospheric pressure and the length of the 3-dimensional linear acceleration vector, the steps per second and the z-value of the gravity sensor passing through the display, and
The figure shows the computed ground truth and a prediction probability value of a neural network based on the \textit{Full Feature Vector}, i.e., a probability value $< 50$\% means that a Wi-Fi connection loss is predicted and vice versa. 
The graphical representation of the sensor values shows that no obvious correlation between one of the sensors and the prediction ground truth exists.
Nevertheless, each of the sensors shows some information that could be useful. 
For example, the atmospheric pressure sensor rises from $t = 100$ to $t = 115$, which could be caused by changing the floor in order to leave the building or by a changing ventilation. In combination with the step counter delta, the first option is more likely, also resulting in a higher likelihood for a Wi-Fi connection loss. Another example is the gravity sensor's z axis that reports about $9.81$ for the time period from $t = 20$ until $t = 35$, which together with the linear acceleration sensor is a good sign for laying flat on a table. This again reduces the likelihood of a Wi-Fi connection loss event.
% ### Ende Beispiel

For the neural network shown on the bottom in Figure~\ref{fig:timo_1518785457}, a 60 seconds observation window has to be filled before the first prediction is performed at $p_{start}$. The classification ends at $p_{end}$, since the operating system reports that Wi-Fi is unavailable.
%As described in Section \ref{sec:design}, 
% The ground truth is computed as a minimum over the prediction window. 
Since Wi-Fi becomes unavailable at $loss$, the ground truth is $0$ from $p_2$ ongoing, matching the 15 seconds prediction window. The neural network classifier matches the ground truth quite well, with the exception of $p_1$, where the classifier predicts the loss slightly too early.
This example shows that the combination of sensors available on today's smartphones can lead to an effective prediction of Wi-Fi connection loss.

\subsection{Machine Learning Results}
This section presents results of training different methods with the data to predict Wi-Fi connection loss: (a) a random forest classifier \cite{liaw2002classification}, and (b) a multi-layer neural network. In particular, we use the MLPClassifier and RandomForest implementations of scikit-learn~\cite{scikit-learn}. 

% \begin{table}[t]
% \centering
% \caption{\textit{Reduced Feature Vector}, random forest (10 trees), randomly split data.}
% \label{results:randomforest}
% \begin{tabular}{r rrr r}
% \toprule
% Event       & Prec.         & Recall        & $F_1$-score   & Support       \\\midrule
% Loss        & \textbf{0.86} & \textbf{0.98} & \textbf{0.91} & 52503         \\
% No Loss     & \textbf{1.00} & \textbf{0.98} & \textbf{0.99} & 438772        \\\midrule
% Total       & \textbf{0.98} & \textbf{0.98} & \textbf{0.98} & 491275        \\\bottomrule
% \end{tabular}
% \end{table}

\paragraph*{Random forest}
Since random forest learning depends on equally distributed samples, the data is down-samp\-led accordingly to match this criterion. 
% Table \ref{results:randomforest} shows the precision, recall, and $F_1$-score of the classification achie\-ved by a 10 tree random forest, learned using the gini criterion \cite{liaw2002classification}. 
The random forest consists of 10 random trees, learned using the Gini criterion.
The overall performance of the random forest is satisfactory, since all values are greater than 0.97. 
However, the precision of the Wi-Fi connection loss class was not very high (0.86), ultimately resulting in triggering early or unnecessary hand\-overs. 

\paragraph*{RSSI-only neural network}
Another basic learning approach is to limit the learner to only use the timeseries of RSSI values, as presented in Section~\ref{sec:rel_work}. During our experiments, different configurations of the neural network were evaluated. 
The overall performance is comparable to the performance presented in the related work. 
The classification quality of the Wi-Fi connection loss class did not exceed an $F_1$-score of 0.95.

\begin{table}[t]
\centering
\caption{\textit{Reduced Feature Vector}, randomly split data, different learners and configurations.}
\label{results:rf-random-overview}
% forest, nn-100, nn-300-200-100, nn-5*400
\begin{tabular}{rrrrr}
\toprule
Metric      & Forest        & \textit{NN 1}          & \textit{NN 2}          & \textit{NN 3}          \\\midrule
Loss Prec.  & 0.89 & 0.95                   & 0.97                   & 0.97                   \\
Loss Recall & 0.98          & 0.94          & 0.95          & 0.95          \\\midrule
$F_1$-score & 0.93          & 0.94                   & 0.96                   & 0.96                   \\\bottomrule
\end{tabular}
\end{table}

\paragraph*{Random Data Split}
The results for neural networks learned with randomly split data depend on the neural network architectures. Table \ref{results:rf-random-overview} provides an overview of different classifier approaches with the \textit{Reduced Feature Vector}. Classifier \textit{NN 1} consists of 100 hidden neurons, \textit{NN 2} of (300, 200, 100) neurons, and \textit{NN 3} of 5 hidden layers containing (400, 400, 400, 400, 400) neurons. All results were achieved using 70\% of the data set exclusively for learning and the remaining 30\% for testing. 
In our experiments, \textit{NN 1} can reach a classification quality comparable to the random forest classifier. The $F_1$-score of the Wi-Fi connection loss class reaches up to 0.94, with either a high precision or a high recall, but never both.
In general, the negative class, representing stable Wi-Fi connections, is predicted well by all tested neural network classifiers. The experiments show that neural networks can reach both high precision and high recall in the positive Wi-Fi connection loss class. 

The results presented in Table~\ref{results:rf-random-overview} show that \textit{NN 2} and \textit{NN 3}  provide reasonably good performance for both precision and recall in the Wi-Fi connection loss class. Even the neural network \textit{NN 2} consisting of three layers shows significant improvements compared to the flat neural network discussed in the previous paragraph. It reaches an $F_1$-score of 0.96 with slightly lower recall or precision.

Other neural network architectures with up to 10 hidden layers were tested. Both precision and recall could not be improved. Splitting the data randomly, \textit{NN 2} and \textit{NN 3} perform equally well and enable a prediction with 97\% precision, 95\% recall, and a combined $F_1$-score of 0.96.

% ToDo: Evaluate NN 2 model generalization. 
%\subsubsection{Model Generalization}
%The generalization abilities of a learned classifier can be evaluated when the classifier is not only applied to previously unseen data, but also to data from a previously unseen device and user. Thus, the data set of \textit{\userSven} was defined as the test set, while the remaining data sets were used for training.

%\begin{table}[t]
%\centering
%\caption{\textit{Reduced Feature Vector}, \textit{NN 3}, test for %\userSven.}
%\label{results:mlp-400-400-400-400-400-sven}
%\begin{tabular}{r rrr r}
% \toprule
% Event           & Prec.         & Recall        & $F_1$-score      & Support       \\\midrule
% Loss        & \textbf{0.92} & \textbf{0.96} & \textbf{0.94} & 25772         \\
% No Loss     & 1.00          & 1.00          & 1.00          & 439487        \\\midrule
% Total     & 0.99          & 0.99          & 0.99          & 465259        \\\bottomrule
% \end{tabular}
% \end{table}

% In Table \ref{results:mlp-400-400-400-400-400-sven}, the classification results of the previously well-performing \textit{NN 3} classifier trained on the \textit{Reduced Feature Vector} is shown. 
\paragraph*{User-based Data Split}
When testing for previously unseen users, the precision of the loss worsens in our prediction. With 0.93, 0.92, and 0.79 precision in the Wi-Fi loss class, the \textit{Reduced Feature Vector} generalizes better compared to the \textit{Full Feature Vector} resulting in 0.91, 0.72, and 0.68 precision.

% When testing on the previously unseen user, the precision of the loss class goes down to 8 \%, while the recall keeps the high 95\%, resulting in a $F_1$-score of 0.91. Even though the previous results are not reached, they show that a reasonably good Wi-Fi connection loss prediction is achieved. The results in this experiment may also be obstructed by the limited availability of different users.

% The user-splitting approach was also evaluated by splitting for \userMsommer and \userTimo. When using the \textit{Full Feature Vector}, the neural network is not able to maintain the high classification quality. The $F_1$-scores for the Wi-Fi loss class reach $0.48$ for \userSven, $0.63$ for \userMsommer and $0.82$ for \userTimo\footnote{The overall $F_1$-score stays at $\sim 0.95$ for all users.}. The same test performed on the \textit{Reduced Feature Vector}, however, results in higher $F_1$-scores: $0.59$ for \userTimo, $0.66$ for \userMsommer and $0.91$ for \userSven. Since the \textit{Reduced Feature Vector} consists of more higher-level and generic features, the classification generalizes better for users.

The results show that the neural networks are capable of generalizing even among different users and devices.
A good classification can be achieved using a neural network with the \textit{Reduced Feature Vector}.
% However incorporating user-specific data allows improvements of precision and recall. 
% When adding the learned neural network to a product, refining a general model using online-learning can be considered. 
Providing a reasonably well basic functionality in the starting phase, with data collected on the device, the classification can be improved during usage.

% \subsubsection{Model Application}
% To further validate the trained neural networks, they were tested in an environment known to us. 
% Figure~\ref{fig:seamless1_scen1_1528536450783} shows the sensor data and prediction of a run performed for the DASH experiments presented in Section~\ref{sec:eval2}. 
For the further model application evaluation, the \textit{Reduced Feature Vector} \textit{NN 3} model was selected.
%,  denoted as \textit{Seamless} in the following analysis.

% \begin{figure}[t]
%     \centering
%     \includegraphics[width=\columnwidth]{figures/seamless1_scen1_1528536450783.pdf}
%     \caption{Sensor data and prediction for a test run.}
%     \label{fig:seamless1_scen1_1528536450783}
% \end{figure}

% The first positive connection loss prediction $p_1$ happens at $t=143$, so that it is 5~seconds early compared to the ground truth that would be 15~seconds before the $loss$ event at $t=163$. The predictions performed after the $loss$ event are not considered and can be skipped, since Wi-Fi connection losses do not need to be predicted when no Wi-Fi is available. Since the classifier is not trained with data after Wi-Fi connection losses, the values after $t=163$ are arbitrary and can be ignored. 

\section{Improved Video Streaming with Seamless MPTCP Handovers}
\label{sec:eval2}
As presented in Section~\ref{sec:eval}, the learned neural network models reliably predict Wi-Fi connection loss with a high precision and recall. To show the usefulness of these results, we evaluated the performance when performing handovers in real-world mobile usage scenarios. In the following, we present a seamless Wi-Fi/cellular handover during DASH video streaming sessions.

\subsection{Seamless Network Connectivity App}
To gather the training data, perform the prediction, and test the applicability of the approach, we implemented a mobile application that performs the following tasks: 

\paragraph*{Sensor data collection and preprocessing}
The sensor readings described in Section~\ref{design:sensors} are cached in memory and written periodically to a local SQLite database on the smartphone. When a run ends, the database is uploaded to a server.
To execute the neural network on the smartphone, the sensor values are preprocessed similarly to the offline learning process. The mean, variance, and the observation window determined offline are used.
\paragraph*{Online prediction}
The offline learned neural networks are transpiled to Java using the sklearn-porter\cite{skpodamo} framework, which allows execution of the same neural networks trained with sklearn on the device. This execution on the Android device allows us to achieve low delays in predictions, independence of Internet access, and protects user privacy. 

\paragraph*{Demonstration \& reporting}
We demonstrate the feasibility of the proposed approach using an embedded DASH video playback functionality. Here, the goal is to highlight potential in improved playback quality and stability made possible by seamless connectivity. 
We use the open movie Elephants Dream,
%\footnote{https://orange.blender.org}
streamed from a server in the university network. The video was pre-encoded using the h.264 encoder for video and AAC for audio, in three bandwidths 1, 2 and 4 MBit/s and a segment length of 2 seconds. 
% representations: 1 Mbit/s (512 x 288), 2 Mbit/s (768 x 432) and 4 Mbit/s (1280 x 720), all using 24 frames per second and a segment length of 2 seconds. 
For video playback, the JavaScript-based DASH.js player (v 2.5.0) was used with a buffer size of 10 seconds in conjunction with the BOLA adaptation algorithm.
To analyze the QoE, we collect and report raw video metrics in each streaming session while the video is playing to the server including stalls, playback bit rates, quality adaptations, and buffer levels for later evaluation.

% \begin{figure}[t]
%     \centering
%     \includegraphics[width=.36\columnwidth]{figures/app_screenshot}
%     \caption{Screenshot of the mobile app showing the DASH video and  Wi-Fi connection loss prediction.}
%     \label{fig:app_screenshot}
% \end{figure}

\paragraph*{MPTCP handovers}
We use the Wi-Fi connection loss prediction to trigger the cellular subflow establishment for MPTCP  \emph{before} the Wi-Fi connection is lost.
We implemented our approach on top of the MPTCP kernel implementation for Android\footnote{https://multipath-tcp.org/pmwiki.php/Users/Android} and the \textit{MultipathControl}\footnote{https://github.com/MPTCP-smartphone-thesis/MultipathControl} app of De Coninck et al. ~\cite{de2016first}.
Furthermore, the video server uses MPTCP version 0.92 with the redundant scheduler and the fullmesh path manager enabled.

\subsection{Experimental Setup}
Our experiments consist of 3 connectivity modes 
in 4 scenarios each performed 5 times, resulting in a total of 60 iterations. The experiments were performed on a Google Nexus 5 smartphone running a rooted Android and the MPTCP Kernel version 0.89.5.
The following connectivity modes were evaluated:
\begin{itemize}
    \item{\textit{Stock} Android:} The default Android mechanism was used to detect Wi-Fi unavailability. During these tests, no transition mechanism was used to have a baseline to compare with.
    \item{\textit{MPTCP}:} To see how MPTCP can improve handover situations, it was enabled for the entire run in these tests. The cellular uplink was used as the second interface, and both client and server used the default scheduler.
    \item{\textit{Seamless}:} During these tests, the \textit{Reduced Feature Vector} neural network in configuration \textit{NN 3} was used, since it showed the most promising results. MPTCP is enabled when a Wi-Fi loss is predicted and disabled when Wi-Fi is available and no loss is predicted for  5 seconds.
\end{itemize}

\begin{figure}[t]
    \centering
    \includegraphics[width=.9\linewidth]{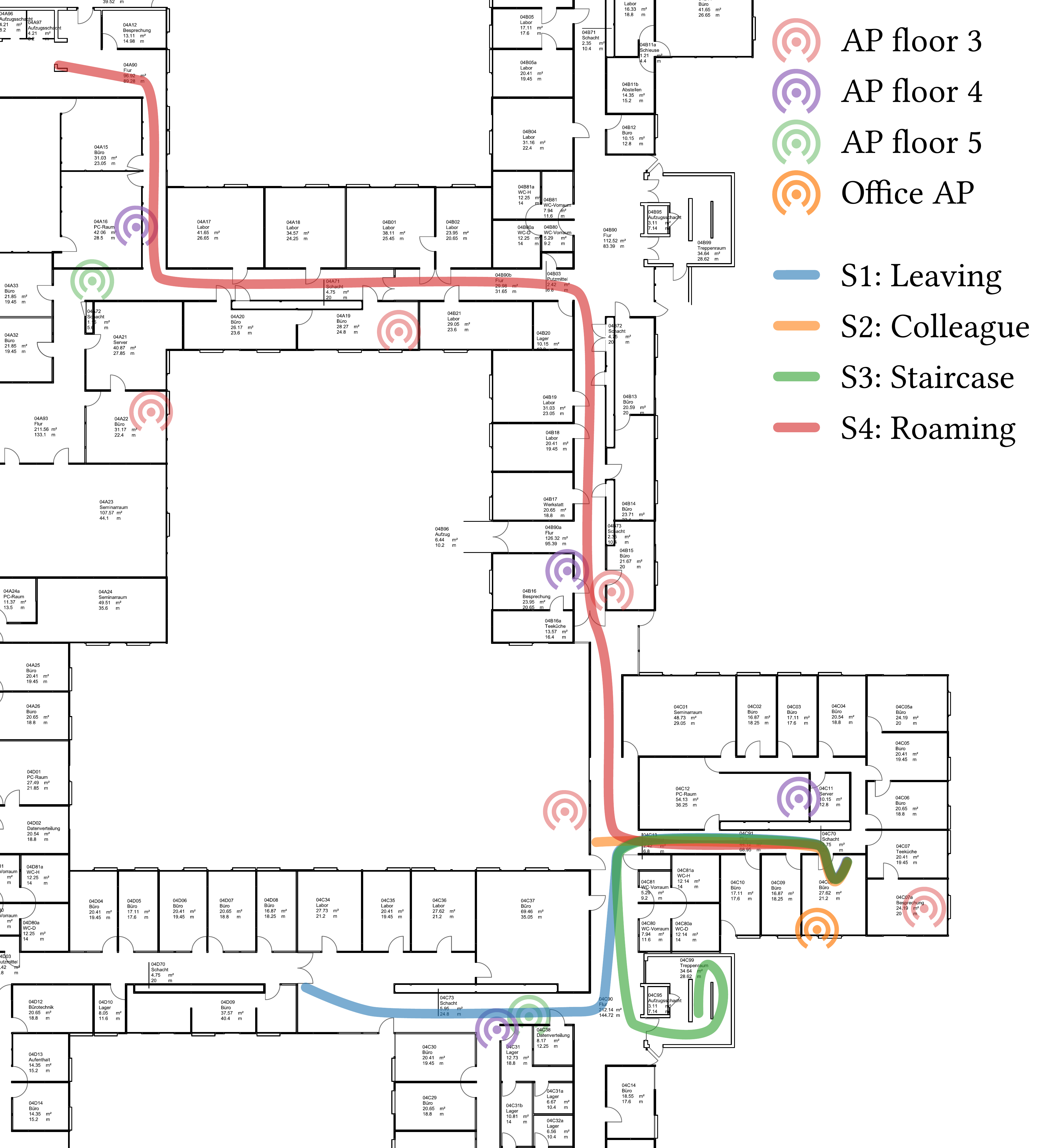}
    \caption{Map with Wi-Fi APs and scenarios routes.}
    \label{fig:experiment_map}
\end{figure}

%\subsubsection{Scenarios}
The following set of routes is chosen to evaluate scenarios in which Wi-Fi connection losses can occur. Figure~\ref{fig:experiment_map} shows the room plan of the university building where the tests were performed.

\paragraph*{Scenario 1: Leaving the office}
Starting in the office, the smartphone is connected to the office Wi-Fi. The tester leaves after 120 seconds of video playback and heads towards the exit of the building. After the Wi-Fi connection is lost (determined in advance, roughly 50 meters) the tester waits for 10 seconds and ends the scenario.  
% The scenario is created to mimic situations when leaving location.
% The Wi-Fi connection degrades and finally is not available anymore. 

\paragraph*{Scenario 2: Visiting a colleague}
The beginning is similar to \textit{Scenario 1}, but the tester walks around about 20 meters away from the office, visiting a colleague, but not leaving the Wi-Fi range. The tester stays for 10 seconds and then walks back to the office. 
% The scenario is created to analyze the behavior in situations where Wi-Fi is still available and usable after a location change. 

\paragraph*{Scenario 3: Using the staircase}
Starting as before, the tester leaves the office on the same route, but then uses the staircase to go up one floor and stays there for 10 seconds.
The scenario shows the impact of a Wi-Fi connection that, while remaining available, is not usable.

\paragraph*{Scenario 4: Wi-Fi roaming support}
Starting in the office, the device is connected to the university network. The tester leaves after 120 seconds and heads towards the other end of the building, roaming between multiple possible Wi-Fi APs  shown in Figure~\ref{fig:experiment_map}. The tester stays near the exit for 10 seconds and then walks back the same route. 
This scenario is created to further investigate the support of roaming gaps in corporate wireless networks where roaming might be available, but is not sufficient to achieve a high QoE.

\subsection{Measuring Quality of Experience}
The DASH video streaming technology is widely available, used by many vendors, and evaluated well. To measure the perceived QoE, several technical values are captured that are used to compute mean opinion scores (MOS)~\cite{stohr2016qoe}, as discussed below.

\subsubsection{Direct Measurements}
During the experiments, the DASH video player reports different technical parameters to a server. At the beginning of a video, the initial buffer has to be filled. This takes time, resulting in \emph{initial stallings} that are perceived to be more disturbing the longer they take. Furthermore, \emph{stalling events} during the video are also reported. 
% Again, less and shorter stalls are perceived as better. 
Apart from the stallings, the \emph{number of adaptations} is counted, since many adaptations also negatively influence the QoE. In addition, the percentage of \emph{time spent in the highest achieved quality} is measured. From a user's perspective, it is better to hold a certain quality as long as possible, even if it is not the best quality available. Since stalling events and quality adaptations partly depend on the buffer level (i.e., how much playable video is in the buffer), the \emph{buffer level} is also captured. The buffer level should be as constant as possible for about 10 seconds.
%These values are reported by the DASH player on the smartphone and characterize the stream played back to the user. 

Finally, a packet dump is performed on the server to allow further analysis of the connections created by MPTCP.
% Since MPTCP uses the standard socket interface, neither the DASH player nor the server know something about the MPTCP state. Therefore, a packet dump is performed on the server to allow further analysis of the connections created by MPTCP.

\subsubsection{QoE Metrics}
\newcommand{\mos}{$MOS_{combined}$\xspace}

Apart from directly evaluating the metrics discussed above, derived metrics are used to capture  relations between these metrics and their impact on QoE. %For evaluating user perception, the MOS-scale is used in literature. 
The $QoE_{stall}$ (Equation (1)) is derived on a MOS scale (where 1 denotes a bad user experience and 5 an excellent one) based on the stalling durations and frequencies during video playback.
Furthermore, $MOS_{quality}$ (Equation (2)) is deduced based on playtime in the highest achieved quality ($t$).
$L$ denotes the average length of all conducted stallings (initially or during video playback) and $N$ the number of stallings, again either initially or during playback. Since our work focuses on Wi-Fi connection loss events, we do not evaluate initial stallings.

\begin{align}
    MOS_{stall} &= 3.5 \times e^{-(0.15 \times L + 0.19) \times N} + 1.5 \\
    MOS_{quality} &= 0.003 \times e^{0.064 \times t \times 100} + 2.498 \\
    MOS_{combined} &= \frac{MOS_{stall} + MOS_{quality}}{2}
\end{align}

Finally, Stohr et al. \cite{stohr2016qoe} propose the average MOS, denoted as \mos (Equation (3)),
%where all mentioned MOS values are combined,
denoting a total user perception not only depending on stalling or quality adaptations. We use \mos to evaluate QoE.

\subsubsection{QoE Experimental Results}
% Since the most recent MPTCP kernel is only available for Android 4.4  (released 2013), a Lenovo Moto Z device with Android 7.1.2 was also used to compare the behavior of a more recent OS with the old version. The results of both versions are comparable, thus the Android 7 findings are not further discussed here.

In Table~\ref{tab:overview}, the overall results of the performed tests are presented, namely the number of stalling events (\# St.) and the average duration of a stalling event ($\varnothing$ St.), the number of adaptations (\# A.), the relative time in the highest playback quality (HQ), and the average transmitted data ($\varnothing$ TD).

\begin{table*}[t]
    \caption{Overview of Experimental Results}
    \label{tab:overview}
    \begin{subtable}{.5\textwidth}
    
        \centering
        \subcaption{Scenario 1: Leaving}
        \label{tab:overview_scen1}
        \begin{tabularx}{0.9\columnwidth}{Xrrrrrr}
        \toprule
        Mode    & \# St. & $\varnothing$ St.  & \# A. & HQ & $\varnothing$ TD \\
        \midrule
        \textit{Stock}   &    3   &        1.46 s       &   23  &  87 \%   &      21.75 MB      \\
        \textit{MPTCP}   &    0   &         0 s         &   20  &  89 \%  &      41.32 MB      \\
        \textit{Seamless} &    0   &         0 s         &   27  &  88 \%  &      36.11 MB      \\
        % \textit{Model} 2 &    0   &         0 s         &   22  &  87 \%  &      27.37 MB      \\
        \bottomrule
        \end{tabularx}
        
    \end{subtable}%
    \begin{subtable}{.5\textwidth}
    
        \centering
        \subcaption{Scenario 2: Colleague}
        \label{tab:overview_scen2}
        \begin{tabularx}{0.9\columnwidth}{Xrrrrrr}
        \toprule
        Mode    & \# St. & $\varnothing$ St.  & \# A. & HQ & $\varnothing$ TD \\
        \midrule
        \textit{Stock}   &    0   &       0 s           &   10  &  92 \%  &      0 MB          \\
        \textit{MPTCP}   &    0   &       0 s           &   10  &  91 \%  &      9.98 MB       \\
        \textit{Seamless} &    0   &       0 s            &   17  &  92 \%  &      9.59 MB       \\
        % \textit{Model 2} &    0   &       0 s           &   10  &  92 \%  &      5.96 MB       \\
        \bottomrule
        \end{tabularx}
    
    \end{subtable}
    \begin{subtable}{.5\textwidth}
    
        \centering
        \subcaption{Scenario 3: Staircase}
        \label{tab:overview_scen3}
        \begin{tabularx}{0.9\columnwidth}{Xrrrrrr}
        \toprule
        Mode    & \# St. & $\varnothing$ St.  & \# A. & HQ & $\varnothing$ TD \\
        \midrule
        \textit{Stock}   &    3   &        2.06 s       &  49   &  80 \%  &      0 MB          \\
        \textit{MPTCP}   &    0   &        0 s          &  32   &  87 \%  &      33.92 MB      \\
        \textit{Seamless} &    0   &        0 s          &  28   &  85 \%  &      16.81 MB      \\
        % \textit{Model 2} &    0   &        0 s          &  30   &  85 \%  &      18.23 MB      \\
        \bottomrule
        \end{tabularx}
        
    \end{subtable}%
    \begin{subtable}{.5\textwidth}
    
        \centering
        \subcaption{Scenario 4: Wi-Fi Roaming}
        \label{tab:overview_scen4}
        \begin{tabularx}{0.9\columnwidth}{Xrrrrrr}
        \toprule
        Mode    & \# St. & $\varnothing$ St.  & \# A. & HQ & $\varnothing$ TD \\
        \midrule
        \textit{Stock}   &    18  &         14.98 s     &  42   &  53 \%  &       0.89 MB      \\
        \textit{MPTCP}   &    0   &         0 s         &  38   &  86 \%  &       71.99 MB     \\
        \textit{Seamless} &    15  &         5.47 s      &  23   &  84 \%  &       15.50 MB     \\
        % \textit{Model 2} &    10  &         9.09 s      &  52   &  57 \%  &       30.1 MB      \\
        \bottomrule
        \end{tabularx}
        
    \end{subtable}
\end{table*}

\paragraph*{Scenario 1}
As shown in Table~\ref{tab:overview_scen1}, the \textit{Stock} tests performed worst with 3 stalling events in total and an average stalling duration of about 1.5 seconds, while neither \textit{MPTCP} nor \textit{Seamless} tests did show any stalling events, which is a significant improvement compared to the stock tests.
%The number of adaptations and playtime at the highest achieved quality are comparable during all tests, not showing any significant differences.
The amount of transferred data over cellular is high in the \textit{MPTCP} test and low in the \textit{Stock} test. \textit{Seamless} results are between these two tests, thus saving cellular data compared to MPTCP, while still avoiding stallings. The results of these tests show that our prediction can avoid the handover gap completely.

% \begin{figure}
%     \begin{subfigure}{\columnwidth}
%         \centering
%         \includegraphics[width=.8\columnwidth, trim={0.1cm 0.25cm 0.1cm 0.25cm}, clip]{bandwidth_plots/stock4_scen1_5}
%         \subcaption{Stock Android}
%         \label{fig:stock4_scen1}
%     \end{subfigure}
%     \begin{subfigure}{\columnwidth}
%         \centering
%         \includegraphics[width=.8\columnwidth, trim={0.1cm 0.25cm 0.1cm 0.25cm}, clip]{bandwidth_plots/seamless1_scen1_4}
%         \subcaption{\textit{Seamless}}
%         \label{fig:seamless1_scen1}
%     \end{subfigure}
%     \caption{\textit{Stock} and \textit{Seamless} in \textit{Scenario 1}.}
% \end{figure}

% Figures~\ref{fig:stock4_scen1} and~\ref{fig:seamless1_scen1} show the bandwidth usage via Wi-Fi and cellular on the left y-axis, and the buffer level of the video on the right y-axis. The x-axis shows the playback time in seconds. The grey shading in the background shows the quality of the video. The darker the color, the better the quality. Finally, a pink background color indicates stalling events. If either Wi-Fi or cellular are not connected to an AP or cell tower, the plot is interrupted at the corresponding points. The plots start at 120 seconds, since the starting phase is identical for all experiments.

% Comparing Figures~\ref{fig:stock4_scen1} and~\ref{fig:seamless1_scen1}, it can be seen, that based on the prediction of \textit{Seamless}, the cellular subflow is established proactively, resulting in a seamless handover and thus no video stalling.

When looking at the buffer levels, video stream quality and the used bandwidth, it can be seen that based on the prediction of \textit{Seamless}, the cellular subflow is established proactively, resulting in a seamless handover and thus no video stalling.

\begin{figure}[t]
    \centering
    \includegraphics[width=.90\columnwidth]{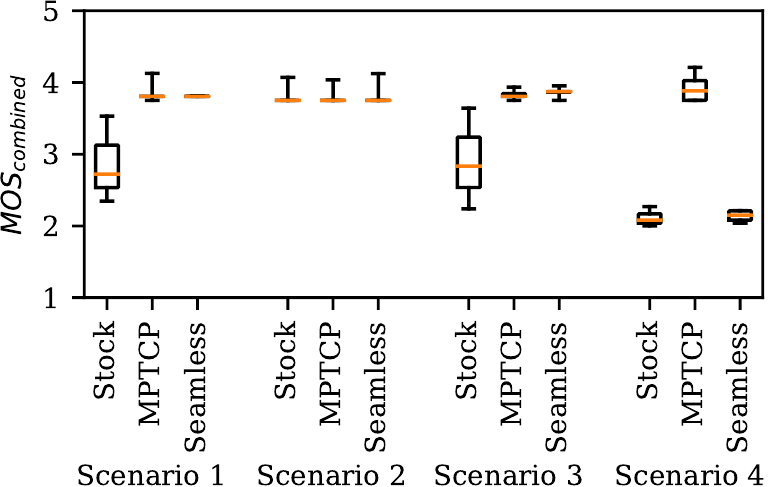}
    \caption{\mos values grouped to connectivity modes and scenarios.}
    \label{fig:qoe_mos}
\end{figure}

Apart from improvements of these technical values, our approach improves QoE for users, as expressed in the \mos. Figure~\ref{fig:qoe_mos} shows the \mos on the y-axis and the different connectivity modes on the x-axis, grouped by scenario. For the stock tests, the \mos is between about 2.5 (poor) and 3.5 (fair), indicating that the playback is not totally unsatisfactory, but far away from a great experience. \textit{Seamless}, on the other hand, achieves a \mos of almost 4, indicating a good QoE, as high as in \textit{MPTCP} tests.

\paragraph*{Scenario 2}
As shown in Table~\ref{tab:overview_scen2}, all tests are comparable for all metrics, showing that our approach does not introduce any negative effects in already good situations. The transferred amount of data over cellular in \textit{Seamless} is about as high as in the \textit{MPTCP} tests. This is because the classifier predicts a Wi-Fi connection loss due to the movement of the smartphone and thus switches to the cellular network, even though this is not necessary. 

Neither the technical metrics like buffer level or used bandwidth, nor the MOS values differ in the these experiments, thus they are not further evaluated here, again indicating that our approach does not worsen the situation by any means.
%\TODO{Arumentation aus der Intro geht auch über Bandbreite / Energie... Muss hier kommentiert werden.}

\paragraph*{Scenario 3}
As shown in Table~\ref{tab:overview_scen3}, the stock tests performed worst with 3 stallings and an average stalling time of about 2 seconds. Additionally, with 49 adaptations and only 80\% of the time at the highest achieved quality, the stock tests perform badly. \textit{MPTCP} and \textit{Seamless} do not stall at all. With 28 and 30 adaptations and 85\% of the time at the highest achieved quality, the results of our approach are as good as in the \textit{MPTCP} tests, again showing significant improvements over the stock implementation. The data usage over cellular shows the same behavior as in \textit{Scenario 1}.
%, again requiring less data compared to MPTCP but more than during stock tests, with the benefit of eliminating stallings during video playback.

\begin{figure}[t]
    \begin{subfigure}{\columnwidth}
        \centering
        \includegraphics[width=.8\columnwidth, trim={0.1cm 0.25cm 0.1cm 0.25cm}, clip]{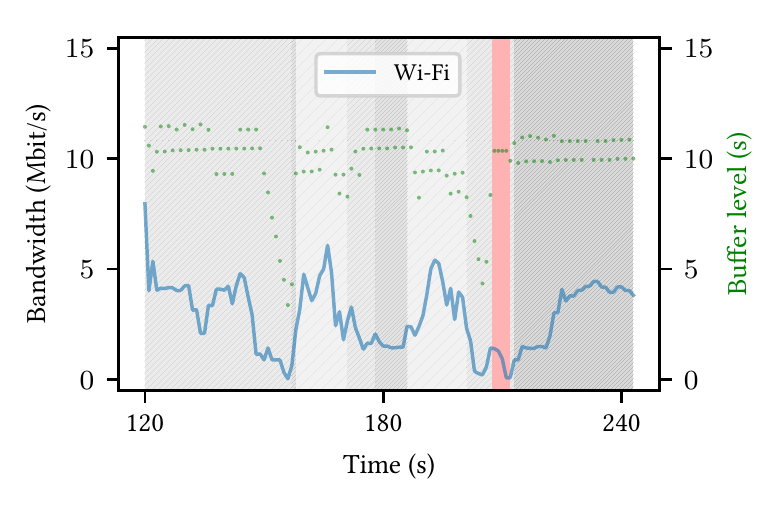}
        \subcaption{Stock Android}
        \label{fig:stock4_scen3}
    \end{subfigure}
    \begin{subfigure}{\columnwidth}
        \centering
        \includegraphics[width=.8\columnwidth, trim={0.1cm 0.25cm 0.1cm 0.25cm}, clip]{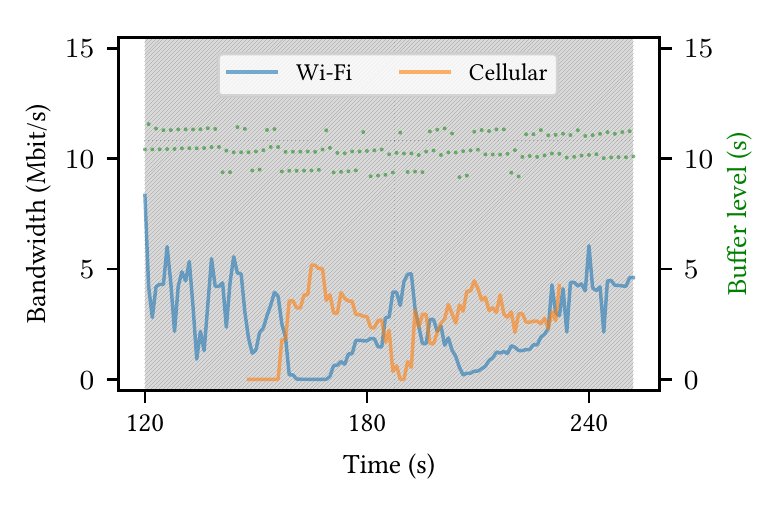}
        \subcaption{Seamless}
        \label{fig:seamless1_scen3}
    \end{subfigure}
\caption{\textit{Stock} and \textit{Seamless} in \textit{Scenario 3}}
\end{figure}

Figures~\ref{fig:stock4_scen3} and~\ref{fig:seamless1_scen3} show bandwidth and buffer level for \textit{Scenario 3}. In the stock tests, the maximum distance is shown in the used bandwidth around seconds 150 and 210. \textit{Seamless} improves this situation and establishes a cellular connection in a timely manner resulting in no stallings.
\mos during the stock tests shows again a relatively bad QoE with about 2.5 to 3.5 compared to the high MOS values of about 4 during MPTCP and tests using our approach.

\paragraph*{Scenario 4}
The results of the stock tests in Table~\ref{tab:overview_scen4} indicate that Android does not handle \textit{Scenario 4} well. The video stalls 18 times and for about 15 seconds on average. The video quality adapts 42 times in total and stays only for 53\% of the time at the highest achieved quality. MPTCP, on the other hand, handles \textit{Scenario 4} very well with no stallings, few quality adaptations, and 86\% of the time at the highest achieved quality.

Although \textit{Seamless} cannot completely cope with the situation, the results are much better than in the stock tests. With 15 stallings and an average stalling duration of 5.5 seconds, just 23 quality adaptations and 84\% percent of the playtime at the highest achievable quality, the results indicate an improved QoE using our approach. The benefits of these improvements come with the cost of using more data (14.61 MB) over the cellular network, but only 21.53\% of cellular data compared to the MPTCP tests.

It is evident that our approach predicts Wi-Fi connection loss correctly, since connection establishment occurs in a timely manner. 
% As shown in Figure~\ref{fig:seamless1_scen4},
However, 
the cellular interface does not reach the high bandwidth used by MPTCP in the same scenario, which might be due to the fact that the cellular interface requires a longer starting phase in the concrete area of the building. 
Also, due to the relatively short Wi-Fi-less gaps investigated in this scenario, the cellular connections are dismantled shortly after they are established. 
A model optimized not only for predicting Wi-Fi connection loss events but also Wi-Fi recovery could improve such scenarios by keeping the cellular link longer alive.

\mos during the \textit{Seamless} and stock tests is comparably bad with a value of about 2, while MPTCP still reaches a MOS of about 4. Nevertheless, our approach reaches a slightly higher QoE than stock Android, as shown in Figure~\ref{fig:qoe_mos}.

% To summarize, the evaluation of our approach shows that predicting Wi-Fi connection loss based on smartphone sensor data does not only work in simulations, but also when it is implemented on real-world devices. 
% Furthermore, our prediction can improve QoE of video streams significantly. 
% Even in challenging scenarios like frequent handovers in a short time period, our approach improves QoE. 
% We demonstrated that our approach is comparable to MPTCP-enabled handover mechanisms regarding QoE, but it can reduce the consumed cellular bandwidth up to 50\%, depending on the scenario. 
% Finally, we also showed that our system does not introduce any negative effects in situations where  Wi-Fi connectivity is not lost.

%\input{4c_overhead}
\section{Conclusion}
\label{sec:conclusion}

In this paper, we proposed a novel data-driven approach to predict Wi-Fi connection loss to perform seamless vertical Wi-Fi/cellular handovers. The approach is based on %a variety of
sensors available in today's smartphones and uses MPTCP to dynamically switch between different wireless connectivity modes. We demonstrated that our trained neural networks reliably predict Wi-Fi connection loss 15 seconds ahead of time when users move around, with a precision of up to 0.97 and a recall of up to 0.98.
Furthermore, we illustrated the benefits of our Wi-Fi connection loss prediction approach with an MPTCP video streaming application. %In particular, 
We showed that our predictions improve the QoE mean opinion score from 2.7 to up to 3.8 for certain scenarios, while reducing the required cellular data usage by up to 50\% compared to traditional MPTCP approaches, with a negligible power consumption overhead.

% \TODO{Alex: wuerde die überschrift hier für future work einfach wegnehmen, inhalt danach mit absatz einfach dahinter. Und auf jeden Fall weniger von der Laenge}

There are several areas for future work.
While the contextual sensors used in our work support high-quality predictions, other more domain-specific sensors might be useful to predict, e.g., Wi-Fi overloads. It would also be interesting to learn predictions
for user/access point combinations. 
To deploy predictors efficiently on off-the-shelf smartphones, lightweight neural networks on dedicated processing engines should be considered. Finally,
%in addition to  Wi-Fi connection loss prediction, 
Wi-Fi connection regain prediction is an interesting area for future research.

% use sensors not avail. in Android API (NEXMON!), 
% per-user/AP model, 
% online learning, 
% lightweight neural nets (maybe on chip/sensorhub)
% Wi-Fi regain prediction...
\section*{Acknowledgment}
%\vspace{-1mm}
This work has been funded by the German Research Foundation (DFG, SFB 1053 MAKI). 
%as part of the project A3, B3, C3 and C5 in the Collaborative Research Center (SFB) 1053 MAKI.

\bibliographystyle{IEEEtranS}
\bibliography{paper_ieee}

% Generated by IEEEtranS.bst, version: 1.12 (2007/01/11)
\begin{thebibliography}{10}
\providecommand{\url}[1]{#1}
\csname url@samestyle\endcsname
\providecommand{\newblock}{\relax}
\providecommand{\bibinfo}[2]{#2}
\providecommand{\BIBentrySTDinterwordspacing}{\spaceskip=0pt\relax}
\providecommand{\BIBentryALTinterwordstretchfactor}{4}
\providecommand{\BIBentryALTinterwordspacing}{\spaceskip=\fontdimen2\font plus
\BIBentryALTinterwordstretchfactor\fontdimen3\font minus
  \fontdimen4\font\relax}
\providecommand{\BIBforeignlanguage}[2]{{%
\expandafter\ifx\csname l@#1\endcsname\relax
\typeout{** WARNING: IEEEtranS.bst: No hyphenation pattern has been}%
\typeout{** loaded for the language `#1'. Using the pattern for}%
\typeout{** the default language instead.}%
\else
\language=\csname l@#1\endcsname
\fi
#2}}
\providecommand{\BIBdecl}{\relax}
\BIBdecl

\bibitem{6587998}
A.~Ahmed, L.~M. Boulahia, and D.~Gaiti, ``Enabling vertical handover decisions
  in heterogeneous wireless networks: A state-of-the-art and a
  classification,'' \emph{IEEE Communications Surveys Tutorials}, vol.~16,
  no.~2, pp. 776--811, 2014.

\bibitem{chen2013measurement}
Y.-C. Chen, Y.-s. Lim, R.~J. Gibbens, E.~M. Nahum, R.~Khalili, and D.~Towsley,
  ``A measurement-based study of multipath {TCP} performance over wireless
  networks,'' in \emph{Internet Measurement Conference}.\hskip 1em plus 0.5em
  minus 0.4em\relax ACM, 2013.

\bibitem{de2016first}
Q.~De~Coninck, M.~Baerts, B.~Hesmans, and O.~Bonaventure, ``A first analysis of
  multipath {TCP} on smartphones,'' in \emph{17th Int. Passive and Active
  Measurements Conference}, vol.~17.\hskip 1em plus 0.5em minus 0.4em\relax
  Springer, 2016.

\bibitem{de2017every}
Q.~De~Coninck and O.~Bonaventure, ``Every millisecond counts: Tuning multipath
  {TCP} for interactive applications on smartphones,'' Technical report.
  Available at http://hdl.handle.net/2078.1/185717, Tech. Rep., 2017.

\bibitem{rfc6824}
A.~Ford, C.~Raiciu, M.~Handley, and O.~Bonaventure, ``{TCP} extensions for
  multipath operation with multiple addresses,'' RFC 6824, Internet Engineering
  Task Force, 2013.

\bibitem{4138149}
S.~Horrich, S.~B. Jamaa, and P.~Godlewski, ``Adaptive vertical mobility
  decision in heterogeneous networks,'' in \emph{3rd Int. Conf. on Wireless and
  Mobile Communications}, March 2007, pp. 44--44.

\bibitem{Khan2017}
M.~Khan, A.~Ahmad, S.~Khalid, S.~H. Ahmed, S.~Jabbar, and J.~Ahmad, ``Fuzzy
  based multi-criteria vertical handover decision modeling in heterogeneous
  wireless networks,'' \emph{Multimedia Tools and Applications}, vol.~76,
  no.~23, pp. 24\,649--24\,674, 2017.

\bibitem{liaw2002classification}
A.~Liaw and M.~Wiener, ``Classification and regression by {RandomForest},''
  \emph{R News}, vol.~2, no.~3, pp. 18--22, 2002.

\bibitem{Lin:2008:NCH:1386109.1386110}
T.~Lin, C.~Wang, and P.-C. Lin, ``A neural-network-based context-aware handoff
  algorithm for multimedia computing,'' \emph{ACM Trans. Multimedia Comput.
  Commun. Appl.}, vol.~4, no.~3, pp. 17:1--17:23, Sep. 2008.

\bibitem{ma2004new}
L.~Ma, F.~Yu, V.~C. Leung, and T.~Randhawa, ``A new method to support
  {UMTS/WLAN} vertical handover using {SCTP},'' \emph{IEEE Wireless
  Communications}, vol.~11, no.~4, pp. 44--51, 2004.

\bibitem{Mansour:2017:SHB:3090354.3090423}
A.~A. Mansour, N.~Enneya, and M.~Ouadou, ``A seamless handover based
  {MIH}-assisted {PMIPV6} in heterogeneous network ({LTE-WIFI}),'' in \emph{2nd
  Int. Conf. on Big Data, Cloud and Applications}.\hskip 1em plus 0.5em minus
  0.4em\relax ACM, 2017, pp. 67:1--67:5.

\bibitem{skpodamo}
\BIBentryALTinterwordspacing
D.~Morawiec, ``sklearn-porter,'' {Transpile} trained Scikit-learn estimators to
  C, Java, JavaScript and others. [Online]. Available:
  \url{https://github.com/nok/sklearn-porter}
\BIBentrySTDinterwordspacing

\bibitem{4289611}
N.~Nasser, S.~Guizani, and E.~Al-Masri, ``Middleware vertical handoff manager:
  A neural network-based solution,'' in \emph{2007 IEEE International
  Conference on Communications}, June 2007, pp. 5671--5676.

\bibitem{Nicholson:2008:BFM:1409944.1409952}
A.~J. Nicholson and B.~D. Noble, ``Breadcrumbs: Forecasting mobile
  connectivity,'' in \emph{14th ACM Int. Conf. on Mobile Computing and
  Networking}, ser. MobiCom '08.\hskip 1em plus 0.5em minus 0.4em\relax ACM,
  2008, pp. 46--57.

\bibitem{PDDRB12}
C.~Paasch, G.~Detal, F.~Duchene, C.~Raiciu, and O.~Bonaventure, ``Exploring
  mobile/wifi handover with multipath {TCP},'' in \emph{ACM SIGCOMM Workshop
  Cellnet}, 2012.

\bibitem{scikit-learn}
\BIBentryALTinterwordspacing
F.~Pedregosa, G.~Varoquaux, A.~Gramfort, V.~Michel, B.~Thirion, O.~Grisel,
  M.~Blondel, P.~Prettenhofer, R.~Weiss, V.~Dubourg \emph{et~al.},
  ``Scikit-learn: Machine learning in python,'' \emph{Journal of Machine
  Learning Research}, vol.~12, no. Oct, pp. 2825--2830, 2011. [Online].
  Available: \url{http://scikit-learn.org/}
\BIBentrySTDinterwordspacing

\bibitem{rfc5944}
C.~E. Perkins, ``{IP} mobility support for {IPv4}, revised,'' RFC 5944,
  Internet Engineering Task Force, 2010.

\bibitem{pluntke2011saving}
C.~Pluntke, L.~Eggert, and N.~Kiukkonen, ``Saving mobile device energy with
  multipath {TCP},'' in \emph{6th International Workshop on MobiArch}.\hskip
  1em plus 0.5em minus 0.4em\relax ACM, 2011, pp. 1--6.

\bibitem{raiciu2012hard}
\BIBentryALTinterwordspacing
C.~Raiciu, C.~Paasch, S.~Barre, A.~Ford, M.~Honda, F.~Duchene, O.~Bonaventure,
  M.~Handley \emph{et~al.}, ``How hard can it be? designing and implementing a
  deployable multipath {TCP},'' in \emph{NSDI}, 2012. [Online]. Available:
  \url{https://dl.acm.org/citation.cfm?id=2228338}
\BIBentrySTDinterwordspacing

\bibitem{stohr2016qoe}
D.~Stohr, A.~Fr{\"o}mmgen, J.~Fornoff, M.~Zink, A.~Buchmann, and W.~Effelsberg,
  ``{QoE} analysis of {DASH} cross-layer dependencies by extensive network
  emulation,'' in \emph{2016 Workshop on QoE-based Analysis and Management of
  Data Communication Networks}.\hskip 1em plus 0.5em minus 0.4em\relax ACM,
  2016, pp. 25--30.

\bibitem{Wanalertlak:2011:BMP:1969330.1969339}
W.~Wanalertlak, B.~Lee, C.~Yu, M.~Kim, S.-M. Park, and W.-T. Kim,
  ``Behavior-based mobility prediction for seamless handoffs in mobile wireless
  networks,'' \emph{Wireless Networks}, vol.~17, no.~3, pp. 645--658, 2011.

\end{thebibliography}

\end{document}